\documentclass[reqno,11pt]{amsart}
\usepackage{amssymb,amsmath}
\usepackage{color}
\usepackage{accents}
\usepackage{texdraw}
\usepackage{eucal}
\usepackage{epic,epsfig}
\usepackage{graphicx}
\usepackage[all]{xy}

\numberwithin{equation}{section}
\DeclareMathAccent{\wtilde}{\mathord}{largesymbols}{"65}
\DeclareMathAccent{\what}{\mathord}{largesymbols}{"62}

\def\m@th{\mathsurround=0pt}
\mathchardef\bracell="0365
\def\upbrall{$\m@th\bracell$}
\def\undertilde#1{\mathop{\vtop{\ialign{##\crcr
    $\hfil\displaystyle{#1}\hfil$\crcr
     \noalign
     {\kern1.5pt\nointerlineskip}
     \upbrall\crcr\noalign{\kern1pt
   }}}}\limits}

\newcommand{\wh}{\widehat}
\newcommand{\wt}{\widetilde}
\newcommand{\ut}{\undertilde}
\def\hypotilde#1#2{\vrule depth #1 pt width 0pt{\smash{{\mathop{#2}
\limits_{\displaystyle\widetilde{}}}}}}
\def\hypohat#1#2{\vrule depth #1 pt width 0pt{\smash{{\mathop{#2}
\limits_{\displaystyle\widehat{}}}}}}

\newcommand{\Ld}{\boldsymbol{\Lambda}}
\newcommand{\tLd}{\,^{t\!}\boldsymbol{\Lambda}}

\newcommand{\bun}{\boldsymbol{1}}
\newcommand{\bOm}{\boldsymbol{\Omega}}
\newcommand{\buk}{\boldsymbol{u}_k}

\newcommand{\tbme}{\,^{t\!}{\boldsymbol{e}}}

\newcommand{\bblu}{\begin{color}{blue}}
\newcommand{\bred}{\begin{color}{red}}
\newcommand{\ecl}{\end{color}}

\newcommand{\bA}{\boldsymbol{A}}

\newcommand{\bC}{\boldsymbol{C}}

\newcommand{\bL}{\boldsymbol{L}}
\newcommand{\bM}{\boldsymbol{M}}

\newcommand{\bO}{\boldsymbol{O}}

\newcommand{\bU}{\boldsymbol{U}}

\newcommand{\gm}{\gamma}

\newcommand{\dd}{\delta}
\newcommand{\sg}{\sigma}

\newcommand{\ld}{\lambda}

\newcommand{\oa}{\omega}
\newcommand{\be}{\begin{equation}}
\newcommand{\ee}{\end{equation}}
\newcommand{\bea}{\begin{eqnarray}}
\newcommand{\eea}{\end{eqnarray}}
\newcommand{\bse}{\begin{subequations}}
\newcommand{\ese}{\end{subequations}}
\newcommand{\nn}{\nonumber}

\newcommand{\bu}{\boldsymbol{u}}
\newcommand{\bv}{{\boldsymbol v}}

\newcommand{\bc}{\boldsymbol{c}}

\newcommand{\bme}{\boldsymbol{e}}

\newcommand{\brr}{\boldsymbol{r}}
\newcommand{\bw}{{\boldsymbol w}}
\newcommand{\mbx}{{\boldsymbol x}}
\newcommand{\mby}{{\boldsymbol y}}
\newcommand{\bz}{{\boldsymbol z}}

\newcommand{\bs}{{\boldsymbol s}}
\newcommand{\btt}{{\boldsymbol t}}

\newcommand{\bphi}{{\boldsymbol \phi}}

\begin{document}
\title[ A higher-rank version of the Q3 Equation]
{A higher-rank version of the Q3 Equation}

\author{Frank W Nijhoff }
\address{
School of Mathematics\\
University of Leeds\\
Leeds LS2 9JT\\
United Kingdom}
\email{nijhoff@maths.leeds.ac.uk}

\begin{abstract}
A lattice system is derived which amounts to a higher-rank analogue of the Q3 equation, the latter 
being an integrable partial difference equation which has appeared in the ABS list of multidimensionally 
consistent quadrilateral lattice equations. By construction this new system incorporates various 
lattice equations of Boussinesq type that were discovered many years ago. A corresponding Lax representation 
is derived. 
\end{abstract}

\maketitle

\section{Introduction}
\setcounter{equation}{0}

In recent years the integrability of partial difference equations (P$\Delta$Es) on the two-dimensional
space-time lattice has become a subject of considerable interest. The early examples of such equations date
back to the 1970s \cite{AL,Hir}, and early 1980s \cite{DJM,NQC,QNCL}, but it is only in the last decade
that their study has moved to the centre-ground of interest in integrable systems. The notion of \textit{multidimensional
consistency}, cf. \cite{NW,BS}, as a key integrability characteristic, has formed the trigger of a wealth of activity in
the subject, leading among others to the celebrated classification result by Adler, Bobenko and Suris (ABS), \cite{ABS},
for scalar quadrilateral P$\Delta$Es.  For the equations in the ABS list (which also comprise some of the quadrilateral
lattice equations already found in the previous era) many results are known, in particular on special solutions, cf.
\cite{AHN,AN,AHN2,NAH,NA,AN2}. A classification result was also formulated recently for three-dimensional scalar octahedral
equations, i.e. equations of lattice KP type, \cite{ABS_KP}, but is markedly still absent for multicomponent and higher-order
lattice equations. In particular it seems that classification results for the ``next'' class of equations beyond the scalar
quadrilateral case, i.e. those of Boussinesq type, requires new ideas and new technical concepts for its treatment.

As far as I am aware, a first example of a lattice Boussinesq (BSQ) equation appeared in \cite{DJM}, this equation being 
a straight dimensional reduction of the Hirota bilinear KP equation, \cite{HM}, it is not clear to me that it is rich enough to 
allow for all appropriate (including intermediate) continuum limits. It is a special case of a more general lattice BSQ
equation which appeared in \cite{GD} where the general class of higher-rank lattice equations of what we coined ``of Gel'fand-Dikii
type'' was considered. (This class, labelled by roots of unity $\oa=\exp(2\pi i/N)$, comprises the KdV type lattice
systems for $N=2$ and BSQ type systems for $N=3$).  The lattice BSQ equation, as given in \cite{GD}, can be presented in the form of 
the following 9-point equation on the two-dimensional lattice:
\begin{eqnarray}
& &\frac{p^3-q^3}{p-q+u_{n+1,m+1}-u_{n+2,m}}\,-\,
\frac{p^3-q^3}{p-q+u_{n,m+2}-u_{n+1,m+1}}   \nn \\
&&\qquad = (p-q+u_{n+1,m+2}-u_{n+2,m+1})(2p+q+u_{n,m+1}-u_{n+2,m+2})  \nn \\
&&\qquad\quad- (p-q+u_{n,m+1}-u_{n+1,m})(2p+q+u_{n,m}-u_{n+2,m+1}) \  . \label{eq:dBSQ}
\end{eqnarray}
in which $u=u_{n,m}$ denotes the dependent variable of the lattice points labelled by $(n,m)\in \mathbb{Z}^2$.
In \eqref{eq:dBSQ} the $p$, $q$ are continuous \textit{lattice} parameters associated with the grid size in the
directions of the lattice given by the independent variables $n$ and $m$ respectively.
For the sake of clarity I prefer to use a notation with lattice shifts denoted by
\[
u=u_{n,m}~\mapsto~\wt{u}=u_{n+1,m}\quad,\quad u=u_{n,m}~\mapsto~\wh{u}=u_{n,m+1}
\]
in terms of which we have also
\[ \wh{\wt{u}}=u_{n+1,m+1}\quad.\quad \wh{\wt{\wt{u}}}=u_{n+2,m+1}\quad,\quad \wh{\wh{\wt{u}}}=u_{n+1,m+2}\quad,\quad\wh{\wh{\wt{\wt{u}}}}=u_{n+2,m+2}\  . \]
Thus, the lattice BSQ equation \eqref{eq:dBSQ} is associated with the following stencil

\vspace{.9cm}

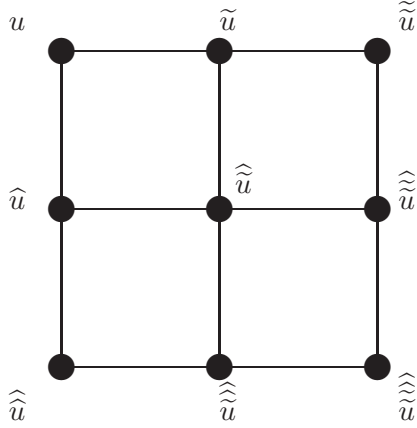
\begin{center}

\setlength{\unitlength}{.7mm}
\begin{picture}(160,60)(-60,0)
\put(0,0){\circle*{5}}
\put(-10,-10){${\large{\wh{\wh{u}}}}$}
\put(0,30){\circle*{5}}
\put(-10,30){${\large{\wh{u}}}$}
\put(0,60){\circle*{5}}
\put(-10,64){${\large{u}}$}
\put(30,0){\circle*{5}}
\put(30,-10){${\large{\wh{\wh{\wt{u}}}}}$}
\put(30,30){\circle*{5}}
\put(33,33){${\large{\wh{\wt{u}}}}$}
\put(30,60){\circle*{5}}
\put(30,64){${\large{\wt{u}}}$}
\put(60,0){\circle*{5}}
\put(64,-10){${\large{\wh{\wh{\wt{\wt{u}}}}}}$}
\put(60,30){\circle*{5}}
\put(64,30){${\large{\wh{\wt{\wt{u}}}}}$}
\put(60,60){\circle*{5}}
\put(64,64){${\large{\wt{\wt{u}}}}$}
\put(0,0){\line(0,1){30}}
\put(0,0){\line(1,0){30}}
\put(30,30){\line(-1,0){30}}
\put(30,30){\line(0,-1){30}}
\put(30,30){\line(1,0){30}}
\put(30,30){\line(0,1){30}}
\put(0,60){\line(1,0){30}}
\put(0,60){\line(0,-1){30}}
\put(60,0){\line(-1,0){30}}
\put(60,0){\line(0,1){30}}
\put(60,60){\line(-1,0){30}}
\put(60,60){\line(0,-1){30}}

\end{picture}

\end{center}
\vspace{.6cm}

In the same paper \cite{GD} we gave a lattice version of the modified BSQ equation, whilst in \cite{N1} the Schwarzian lattice BSQ was presented.
All three equations are specialisations of a more general lattice BSQ type equation, which was presented in \cite{N1,DIGP} and which can be cast
into the following form:
\bea\label{eq:interBSQ}
&& \frac{ \left[ P_a P_b (u \wh{u}+\wt{u}\wh{\wt{u}}) - Q_a Q_b(u\wt{u}+\wh{u}\wh{\wt{u}})-(p^3-q^3) (\wh{u}\wt{u}+ u \wh{\wt{u}})\right]\!\wh{\phantom{a}} }
{\left[ P_a P_b (u \wh{u}+\wt{u}\wh{\wt{u}}) - Q_a Q_b(u\wt{u}+\wh{u}\wh{\wt{u}})-(p^3-q^3) (\wh{u}\wt{u}+ u \wh{\wt{u}})\right]\!\wt{\phantom{a}} }= \nn \\
&&\qquad  = \frac{(Q_a u-Q_b\wh{u})\,(P_a \wh{\wh{\wt{u}}}-P_b \wh{\wh{\wt{\wt{u}}}})\,(Q_aP_b\wh{\wt{u}}-P_a Q_b\wh{\wh{u}})}
{(P_a u-P_b\wt{u})\,(Q_a \wh{\wt{\wt{u}}}-Q_b \wh{\wh{\wt{\wt{u}}}})\,(Q_aP_b\wt{\wt{u}}-P_a Q_b\wh{\wt{u}})}
\eea
In eq. \eqref{eq:interBSQ} the $P_a$, $P_b$, $Q_a$ and $Q_b$ are given by
\[ P_a=\sqrt{p^3-a^3}\quad, \quad P_b=\sqrt{p^3-b^3} \quad, \quad Q_a=\sqrt{q^3-a^3} \quad, \quad Q_b=\sqrt{q^3-b^3}\  , \]
where $a$ and $b$ are some additional (continuous) parameters. If $b=a$ \eqref{eq:interBSQ} reduces to the lattice Schwarzian BSQ equation, whilst in
the limit $b\rightarrow \infty$, $b u\rightarrow v$, we obtain the modified BSQ for the variable $v$. In the double limit $a,b\rightarrow\infty$ we recover
\eqref{eq:dBSQ} after an appropriate scaling. Thus, \eqref{eq:interBSQ} can be viewed as an equation interpolating between these various cases, in the
same way as the so-called NQC equation of \cite{NQC} interpolates between various lattice equations of KdV type.

The mentioned lattice BSQ equations can also be written in multi-component form, and as was shown in \cite{DIGP} the modified BSQ can be written as 2-component
system of P$\Delta$Es on an elementary quadrilateral, and as such multidimensional consistency in the sense of \cite{NW,BS} can be easily shown for that
system. In \cite{W,TN2} eq. \eqref{eq:dBSQ} was written as a three-component system on the elementary quadrilateral, whilst the multidimensional consistency
for the 9-point scheme of the scalar form was studied in \cite{W}. Some further multicomponent BSQ type systems, generalizing the 3-component version of the
lattice BSQ, were given in a recent paper \cite{H}. We mention also that explicit soliton solutions of the lattice BSQ equation were constructed in \cite{HZ,MK},
although it must be said that soliton solutions can also quite easily be inferred from the direct linearization scheme presented in \cite{GD,W}.

In the present paper I construct what I believe is a novel lattice system which incorporates the previously known BSQ type equations, including the 
modified and Schwarzian BSQ equations and \eqref{eq:interBSQ}. The construction of this system is based on the structure of the Q3 solutions, which is a pivotal 
equation in the ABS list, \cite{ABS}. In fact it was shown in \cite{NAH}, cf. also \cite{AHN2,NA}, that solutions of Q3 
can be written as a linear combination of four solutions of the NQC equation of \cite{NQC}, which is the analogue of \eqref{eq:interBSQ} in the quadrilateral scalar case.
Here I perform a similar construction for the higher-rank case, leading in this case to a non-autonomous lattice system which I believe constitutes the
proper analogue of the Q3 equation. Since, at the present moment, even a preliminary classification of higher-rank or multicomponent lattice equations is lacking,
I believe these results could form a first step towards an insight into the structures behind lattice equations in the higher-rank case. In section 2 I will present 
the basic construction of solutions for Q3, which motivates the generalization to the BSQ case. The latter case is markedly more complicated, and in section 3
the system of basic constitutive relations is derived, from which in section 4 I derive the relations forming the system. 
Multidimensional consistency is built into the system by the entire construction, but the integrability is made explicit by presenting in section 5 a Lax representation. 
The whole construction can, in principle, be generalized naturally to the whole Gel'fand-Dikii hierarchy, of which the BSQ system represents the $N=3$ case.

\section{Constitutive system}
\setcounter{equation}{0}

\subsection{Infinite matrix scheme}

The starting point of the construction is the following integral
\be\label{eq:bC}
\bC= \sum_{j=1}^N \int_{\Gamma_j}\,d\ld_j(k)\,\rho_k \bc_k\,\bc^t_{-\oa^jk}\sg_{-\oa^jk}\  ,
\ee
over a yet unspecified set of contours or arcs $\Gamma_j$ in the complex plane of a variable $k$, and in which for  $N\geq 2$ a positive integer,
$\oa=\exp(2\pi i/N)$ is the $N^{\rm th}$ root of unity. The factors $\rho_k$, $\sg_{k'}$ are discrete exponential functions of
$k$, $k'$ respectively, given by
\be\label{eq:rhosgk}
\rho_k(n,m)= (p+k)^n (q+k)^m\rho_k(0,0)\quad,\quad \sg_{k'}(n,m)= (p-k')^{-n} (q-k')^m\sg_{k'}(0,0)\  ,
\ee
and the infinite-component vector $\bc_k=(k^j)_{j\in\mathbb{Z}}$ denotes the vector of a basis of monomials in the variable $k$.
The integration measures $d\ld_j(k)$ remain unspecified, but we assume that basic operations (such as differentiations w.r.t. parameters
or applying shifts in the variables $n$ and $m$) commute with the integrations.

In accordance with the notation introduced in section 1, we denote shifts over one unit in the variables $n$, and $m$ respectively by
the operations $\wt{\phantom{a}}$ and $\wh{\phantom{a}}$, which implies for the factors $\rho_k$, $\sg_{k'}$
\[ \wt{\rho}_k=\rho_k(n+1,m)=(p+k)\rho_k\quad,\quad  \wh{\rho}_k=\rho_k(n,m+1)=(q+k)\rho_k\ , \]
and
\[ \wt{\sg}_{k'}=\sg_{k'}(n+1,m)=(p-k')^{-1}\sg_{k'}\quad,\quad  \wh{\sg}_{k'}=\sg_{k'}(n,m+1)=(q-k')^{-1}\sg_{k'}\ . \]
These relations imply the following linear relations for the matrix $\bC$
\be\label{eq:bCdyn}
\wt{\bC}\,(p-\tLd)=(p+\Ld)\,\bC\quad,\quad \wh{\bC}\,(q-\tLd)=(q+\Ld)\,\bC\  ,
\ee
where we have introduced the matrices $\Ld$ and $\tLd$ which are defined by their actions on the vector $\bc$, and on its transposed vector, as follows:
\[ \Ld\,\bc_k=k\,\bc\quad,\quad \bc^t_{k'}\tLd=k'\,\bc^t_{k'}\  . \]

The following ingredients determine the structure:\\
\textit{i)} A Cauchy matrix $\bOm$ obeying the relation ~$\bOm\Ld+\tLd\bOm=\bO$~,
where $\bO$ is a rank 1 projection matrix, obeying $\bO^2=\bO$. \\
\textit{ii)}  A matrix ~$\bU$~ obeying the relation ~$\bU=\bC-\bU\,\bOm\,\bC$~ \\
\textit{iii)} A vector ~$\bu_k$~ defined by ~$\bu_k=\rho_k(\bc_k-\bU\,\bOm\,\bc_k)$~\\

\noindent In terms of these objects the following sets of relations involving the shift $\wt{\phantom{a}}$ can be derived:
\bse\label{eq:NUrels}\bea
\wt{\bU}\,(p-\tLd)&=& (p+\Ld)\bU-\wt{\bU}\,\bO\,\bU\  , \label{eq:NUrels_a} \\
\bU\,\left[ \prod_{j=1}^{N-1}(\oa^j p-\tLd)\right] &=& \left[\prod_{j=1}^{N-1}(\oa^j p+\Ld)\right]\wt{\bU} -\bU\,
\sum_{j=0}^{N-2} \left[ \prod_{k=1}^j (\oa^kp-\tLd)\right]\bO \left[\prod_{k=j+2}^{N-1} (\oa^kp+\Ld)\right]\,\wt{\bU}\  , \nn \\
&&  \label{eq:NUrels_b} \\
\bU\,(-\tLd)^N &=& \Ld^N\,\bU -\bU\,\sum_{j=0}^{N-1} (-\tLd)^j\bO\Ld^{N-1-j}\,\bU\  , \label{eq:NUrels_c}
\eea\ese
(see \cite{GD}). By virtue of the covariance of the dynamics in terms of the variables $n$ and $m$, similar relations to \eqref{eq:NUrels_a}
and \eqref{eq:NUrels_b} with the shift $\wt{\phantom{a}}$ replaced by $\wh{\phantom{a}}$ while replacing $p$ by $q$. The associated linear
problems (Lax pairs) are derived in terms of the object $\bu_k$, for which we have the following set of constitutive relation, \cite{GD},
\bse\label{eq:Nukrels}\bea
\wt{\bu}_k &=& (p+\Ld)\bu_k-\wt{\bU}\,\bO\,\bu_k\  , \label{eq:Nukrels_a} \\
(k^N-(-p)^N)\bu_k &=& \left[\prod_{j=1}^{N-1}(\oa p+\Ld)\right]\wt{\bu}_k -\bU\,
\sum_{j=0}^{N-2} \left[ \prod_{k=1}^j (\oa^kp-\tLd)\right]\bO \left[\prod_{k=j+2}^{N-1} (\oa^kp+\Ld)\right]\,\wt{\bu}_k \  , \nn \\
&& \label{eq:Nukrels_b} \\
k^N\bu_k &=& \Ld^N\,\bu_k -\bU\,\sum_{j=0}^{N-1} (-\tLd)^j\bO\Ld^{N-1-j}\,\bu_k\  , \label{eq:Nukrels_c}
\eea\ese
and similar relations to \eqref{eq:ukrels_a} and \eqref{eq:ukrels_b} with the shift $\wt{\phantom{a}}$ replaced by $\wh{\phantom{a}}$ while replacing $p$ by $q$.

For any fixed value of $N$, these abstract equations form an infinite set of recurrence relations defining the dynamics in terms of the independent variables $n$,
$m$ on the objects $\bU$ and $\buk$ taking values in an abstract vector space $\mathcal{V}$, and an adjoint vector $\tbme$ in its dual $\mathcal{V}^{\ast}$.
To give a concrete realisation, we can choose a fixed vector $\bme$ and use the matrices $\Ld$ and $\tLd$ to define a natural gradation in this vector space,
In terms these, we realize quantities such as $\bC$ and $\bU$ as infinite-dimensional matrices. In fact, setting
 \[ U_{i,j}:= \tbme\tLd^i\bU\Ld^j\bme\quad,\quad i,j\in\mathbb{Z}\   , \]
we obtain infinite, i.e., $\mathbb{Z}\times\mathbb{Z}$ matrices, on which $\Ld$ and $\tLd$ act as index-raising operators acting from the left and from the right.
In this realization the matrix ~$\bO=\bme\,\tbme$~ is the projector defined by these vectors, and this defines the "central" entry in the space of
$\mathbb{Z}\times\mathbb{Z}$ matrices. We will define concrete objects in terms of the matrix $\bU$, where we allude to this realization.

\subsection{Basic objects and their relations: KdV case $(N=2)$}
Thus, for the case $N=2$ the relations \eqref{eq:NUrels} take the form:
\bse\label{eq:2Urels}\bea
\wt{\bU}\,(p-\tLd)&=& (p+\Ld)\bU-\wt{\bU}\,\bO\,\bU\  , \label{eq:2Urels_a} \\
\bU\,(p+\tLd)&=& (p-\Ld)\wt{\bU} +\bU\,\bO\,\wt{\bU} \  , \label{eq:2Urels_b} \\
\bU\,\tLd^2 &=& \Ld^2\,\bU -\bU\,\left( \bO\,\Ld-\tLd\,\bO\right)\,\bU\  , \label{eq:2Urels_c}
\eea\ese
whereas \eqref{eq:Nukrels} become
\bse\label{eq:ukrels}\bea
\wt{\bu}_k &=& (p+\Ld)\bu_k-\wt{\bU}\,\bO\,\bu_k\  , \label{eq:Nukrels_a} \\
(p^2-k^2)\bu_k &=& (p-\Ld)\wt{\bu}_k +\bU\,\bO\,\wt{\bu}_k \  , \label{eq:ukrels_b} \\
k^2\bu_k &=& \Ld^2\,\bu_k -\bU\,\left( \bO\,\Ld-\tLd\,\bO\right)\,\bu_k\  , \label{eq:ukrels_c}
\eea\ese
Since \eqref{eq:2Urels_b} can be obtained from \eqref{eq:2Urels_a} by matrux transposition, the system admits solutions that are symmetric
with regard to this operation, i.e., $\bU^t=\bU$.

We shall briefly show how the Q3 equation, \cite{ABS}, arises in a simple manner, presenting the essence of the derivation in \cite{NAH}.
In fact, we need to introduce the following quantities:
\bse\label{eq:objsKdV}\begin{eqnarray}
&& v_a:= 1-\tbme\,(a+\Ld)^{-1}\,\bU\,\bme =1-\tbme\,\bU\,(a+\tLd)^{-1}\,\bme\  , \\
&& s_a:= a-\tbme\,(a+\Ld)^{-1}\,\bU\,\tLd\,\tbme =a-\tbme\,\Ld\,\bU\,(a+\tLd)^{-1}\,\bme\  .
\end{eqnarray}\ese
as well as the quantity $u_0:=U_{0,0}$ and
\be\label{eq:KdVs}
s_{a,b}:=\tbme\,(a+\Ld)^{-1}\,\bU\,(b+\tLd)^{-1}\bme=s_{b,a}
\ee
Here $a$ and $b$ are arbitrary scalar parameters, and we have suppressed the multiplication by a unit
matrix in expressions of the type $a+\Ld:=a\bun+\Ld$.
The following relations follow from \eqref{eq:2Urels_a} and \eqref{eq:2Urels_b}:
\bse\label{eq:svrels}\bea
\wt{s}_a &=& (p+u_0)\wt{v}_a-(p-a) v_a \  , \\
s_a &=& (p+a)\wt{v}_a-(p-\wt{u}_0)v_a\
\eea\ese
as well as the relation
\be\label{eq:KdVsrel}
1+(p-a) s_{a,b}-(p+b)\wt{s}_{a,b}=\wt{v}_a v_b\
\ee
and using the symmetry of $s_{a,b}$ we have another relation of this type with $a$ and $b$ interchanged.

Introducing the plane wave factors
$$ \rho_a(n,m)=\left(\frac{p+a}{p-a}\right)^n\left(\frac{q+a}{q-a}\right)^m\rho_a(0,0)\quad,\quad
\rho_b(n,m)=\left(\frac{p+b}{p-b}\right)^n\left(\frac{q+b}{q-b}\right)^m\rho_a(0,0)\  , $$
we can now introduce the two-component vectors:
\be\label{eq:KdVvecs}
\bv_a:= \left(\begin{array}{c} \rho_a^{1/2} v_a \\ \rho_{-a}^{1/2} v_{-a}\end{array}\right)\quad , \quad
\bs_a:= \left(\begin{array}{c} \rho_a^{1/2} s_a \\ \rho_{-a}^{1/2} s_{-a}\end{array}\right)
\ee
and the quantities:
\bse\label{eq:KdVuU}\bea
&& u:= \left(\rho^{1/2}_a\,,\,\rho^{1/2}_{-a}\right)\left( \begin{array}{cc} A S_{a,b} & BS_{a,-b} \\
C S_{-a,b} & DS_{-a,-b}\end{array} \right) \left(\begin{array}{c} \rho_b^{1/2} \\ \rho_{-b}^{1/2}\end{array}\right)\  , \\
&& U:= \bv_a^t\,\bA\,\bv_b\   .
\eea\ese
where $S_{a,b}=:s_{a,b}-1/(a+b)$,  and where $\bA$ is the matrix:
\[ \bA=\left(\begin{array}{cc} A & B \\ C & D \end{array}\right)\  . \]

The relations \eqref{eq:svrels} lead to the vector relations
\bse\label{eq:sv_vecrels}\bea
\wt{\bs}_a &=& (p+u_0)\wt{\bv}_a-P_a v_a \  , \\
\bs_a &=& P_a\wt{\bv}_a-(p-\wt{u}_0)\bv_a\
\eea\ese
and similar relations for the dynamics in the variable $m$, with the usual interchangements and changing $P_a$ into
$Q_a$, where
\[ P_a=\sqrt{p^2-a^2}\quad,\quad Q_a=\sqrt{q^2-a^2}\  , \]
and we will also introduce the quantities ~$P=P_aP_b$~, ~$Q=Q_aQ_b$~.

The relation \eqref{eq:KdVsrel} and its counterpart interchanging $a$ and $b$, now become
\be\label{eq:uurels}
P_a u-P_b \wt{u} =\wt{\bv}_a^t \bA\bv_b\quad P_b u-P_a\wt{u}=\bv_a^t \bA \wt{\bv}_b\  ,
\ee
where it should be noted that the matrix $\bA$ need \textit{not} be symmetric.

The construction of the Q3 equations from these data now follows the following steps:
\paragraph{i) Biquadratic identity:} Consider the following 2$\times$2 matrix
\begin{equation}\label{eq:Edet}
E_{a,b}=\left( \begin{array}{ccc}\bv_a^t\bA\bv_b &,& \bv_a^t\bA\bs_b \\ \bs_a^t\bA\bv_b &,& \bs_a^t\bA\bs_b
\end{array}\right)\   .
\end{equation}
The determinant of this matrix can be computed in two different ways; on the one had we have, using \eqref{eq:sv_vecrels} the sequel:
\begin{eqnarray*}
\det(E_{a,b})&=& P_b\left| \begin{array}{ccc}\bv_a^t\bA\bv_b &,& \bv_a\bA\wt{\bv}_b \\  \bs_a^t\bA\bv_b &,& \bs_a\bA\wt{\bv}_b
\end{array}\right|=P_aP_b \left| \begin{array}{ccc}\bv_a^t\bA\bv_b &,& \bv_a\bA\wt{\bv}_b \\  \wt{\bv}_a^t\bA\bv_b &,& \wt{\bv}_a\bA\wt{\bv}_b
\end{array}\right| \\
&=& \left| \begin{array}{ccc} U &,& P_bu-P_a\wt{u} \\ P_au-P_b\wt{u} &,& \wt{U} \end{array}\right|=
P\left[ U\wt{U}-P(u^2+\wt{u}^2)+(2p^2-a^2-b^2)u\wt{u}\right]\   .
\end{eqnarray*}
On the other hand,using a special version of the Weinstein-Oranszajn formula, the determinant equals
$$ \det(E_{a,b})=\det(\bA)\,\det(\bv_a\,,\,\bs_a)\,\det(\bv_b\,,\,\bs_b)=\alpha\beta\,\det(\bA)\  , $$
where $\alpha:=\det(\bv_a\,,\,\bs_a)$ and $\beta:=\det(\bv_b\,,\,\bs_b)$ are constants, as can be seen from
the chain:
$$ \wt{\alpha}=\det(\wt{\bv}_a\,,\,\wt{\bs}_a)=-P_a\det(\wt{\bv}_a\,,\,\bv_a)=\det(\bv_a\,,\,P_a\wt{\bv}_a)=\det(\bv_a\,,\,\bs_a)=\alpha\  , $$
using once again the vector relations \eqref{eq:sv_vecrels}, and a similarly for $\beta$, and the same argument holds for the shift in the
variable $m$.

Thus, equating both sides of the evaluation of $\det(E_{a,b})$ we obtain the factorization of the biquadratic
\be\label{eq:biquadfact}
U\wt{U}=P(u^2+\wt{u}^2)-(2p^2-a^2-b^2)u\wt{u}+\frac{\alpha\beta}{P}\det(\bA)\  ,
\ee

\paragraph{ii) Q3 equation:} To obtain the main equation of interest, Q3, for the variable $u$ we proceed as follows.
Consider the relations \eqref{eq:uurels} and insert them into the following determinant
\begin{eqnarray*}
&& \left| \begin{array}{ccc}\bv_a^t\bA\wt{\bv}_b &,& \bv_a^t\bA\wh{\bv}_b \\  \wh{\wt{\bv}}_a^t\bA\wt{\bv}_b &,& \wh{\wt{\bv}}_a^t\bA\wh{\bv}_b
\end{array}\right|= \left| \begin{array}{ccc} P_bu-P_a\wt{u} &,& Q_b u-Q_a\wh{u} \\ Q_a \wt{u}-Q_b\wh{\wt{u}} &,&
P_a\wh{u}-P_b\wh{\wt{u}} \end{array}\right| = \\
&& =P(u\wh{u}+\wt{u}\wh{\wt{u}})-Q(u\wt{u}+\wh{u}\wh{\wt{u}})-(p^2-q^2)(\wt{u}\wh{u}+u\wh{\wt{u}})
\end{eqnarray*}
On the other hand we have that the starting determinant equals:
\begin{eqnarray*}
&& \frac{1}{Q_b}\left| \begin{array}{ccc}\bv_a^t\bA\wt{\bv}_b &,& \bv_a^t\bA\left[P_b\wt{\bv}_b-(p-q+\wh{u}_0-\wt{u}_0)\bv_b\right] \\
\wh{\wt{\bv}}_a^t\bA\wt{\bv}_b &,& \wh{\wt{\bv}}_a^t\bA\left[P_b\wt{\bv}_b-(p-q+\wh{u}_0-\wt{u}_0)\bv_b\right]
\end{array}\right| \\
&& = -\frac{p-q+\wh{u}_0-\wt{u}_0}{P_bQ_b}\left| \begin{array}{ccc} \bv_a^t\bA\bs_a &,& \bv_a^t\bA \bv_b \\
\wh{\wt{\bv}}_a^t\bA\bs_b &,& \wh{\wt{\bv}}_a^t\bA\bv_b \end{array}\right|  \\
&&= -\frac{p-q+\wh{u}_0-\wt{u}_0}{P_bQ_bQ_a}\left| \begin{array}{ccc} \bv_a^t\bA\bs_a &,& \bv_a^t\bA \bv_b \\
\left(-P_a\bv_a^t+(p+q+u_0-\wh{\wt{u}}_0)\wt{\bv}_a^t\right)\bA\bs_b &,& \left(-P_a\bv_a^t+(p+q+u_0-\wh{\wt{u}}_0)\wt{\bv}_a^t\right)\bA\bv_b  \end{array}\right| = \\
&& =-\frac{(p-q+\wh{u}_0-\wt{u}_0)(p+q+u_0-\wh{\wt{u}}_0)}{P_aP_bQ_aQ_b} \left| \begin{array}{ccc}\bv_a^t\bA\bs_b &,& \bv_a^t\bA\bv_b \\ \bs_a^t\bA\bs_b &,& \bs_a\bA\bv_b
\end{array}\right|=\frac{p^2-q^2}{PQ}\alpha\beta\det(\bA)\   .
\end{eqnarray*}
Here we have made use of the intermediate relations
\begin{eqnarray*}
&& P_a\wt{\bv}_a-Q_a\wh{\bv}_a=(p-q+\wh{u}_0-\wt{u}_0)\bv_a \  , \\
&& P_a\bv_a+Q_a\wh{\wt{\bv}}_a=(p+q+u_0-\wh{\wt{u}}_0)\wt{\bv}_a \  ,
\end{eqnarray*}
as well as of the following equation for $u_0$
\be\label{eq:H1}
(p-q+\wh{u}_0-\wt{u}_0)(p+q+u_0-\wh{\wt{u}}_0)=p^2-q^2\  ,
\ee
which is the H1 equation (lattice potential KdV equation) in the terminology of the paper \cite{ABS}.
Thus, we obtain in the case $N=2$ the Q3 equation from a simple structure of determinants. We will now follow the same path
for the case $N=3$, leading to a higher-rank case, but we will see that the structure is considerably more involved in this case.

\section{Higher rank case}
\setcounter{equation}{0}

\subsection{Basic objects and their relations: BSQ case $(N=3)$}

In the case $N=3$ the set of relations

Thus, for the case $N=3$ the relations \eqref{eq:NUrels} take the form:
\bse\label{eq:Urels}\bea
\wt{\bU}\,(p-\tLd)&=& (p+\Ld)\bU-\wt{\bU}\,\bO\,\bU\  , \label{eq:Urels_a} \\
\bU\,(\oa p-\tLd)(\oa^2 p-\tLd) &=& (\oa p+\Ld)(\oa^2 p+\Ld)\wt{\bU} +\bU\,\left[ p\bO-(\bO\,\Ld-\tLd\,\bO)\right]\,\wt{\bU} \  , \label{eq:Urels_b} \\
\bU\,(-\tLd)^3 &=& \Ld^3\,\bU -\bU\,\left( \bO\,\Ld^2-\tLd\,\bO\,\Ld+\tLd^2\,\bO \right)\,\bU\  , \label{eq:Urels_c}
\eea\ese
and similar relations to \eqref{eq:NUrels_a} and \eqref{eq:NUrels_b} with the shift $\wt{\phantom{a}}$ replaced by $\wh{\phantom{a}}$ while replacing $p$ by $q$,
as well as
\bse\label{eq:ukrels}\bea
\wt{\bu}_k &=& (p+\Ld)\bu_k-\wt{\bU}\,\bO\,\bu_k\  , \label{eq:ukrels_a} \\
(p^3+k^3)\bu_k &=& (\oa p+\Ld)(\oa^2 p+\Ld)\wt{\bu}_k +\bU\,\left[ p\bO-(\bO\,\Ld-\tLd\,\bO)\right]\,\wt{\bu}_k \  , \label{eq:ukrels_b} \\
k^3\bu_k &=& \Ld^3\,\bu_k -\bU\,\left( \bO\,\Ld^2-\tLd\,\bO\,\Ld+\tLd^2\,\bO \right)\,\bu_k\  , \label{eq:ukrels_c}
\eea\ese
(and similar relations to \eqref{eq:ukrels_a} and \eqref{eq:ukrels_b} with the shift $\wt{\phantom{a}}$ replaced by $\wh{\phantom{a}}$ while replacing $p$ by $q$).
For the sake of obtaining from the constitutive relations \eqref{eq:Urels} closed-form equations,
we introduce the following objects:
\bse\label{eq:objs}\bea
v_a:= 1-\tbme\,(a+\Ld)^{-1}\,\bU\,\bme\quad,\quad w_b:=1+\tbme\,\bU\,(b-\tLd)^{-1}\,\bme\  , \\
s_a:= a-\tbme\,(a+\Ld)^{-1}\,\bU\,\tLd\,\bme\quad,\quad t_b:=-b+\tbme\,\Ld\,\bU\,(b-\tLd)^{-1}\,\bme\  , \\
r_a:= a^2-\tbme\,(a+\Ld)^{-1}\,\bU\,\tLd^2\,\bme\quad,\quad z_b:=b^2+\tbme\,\Ld^2\,\bU\,(b-\tLd)^{-1}\,\bme\  ,
\eea\ese
but in particular the following object:
\begin{equation}\label{eq:s}
 s_{a,b}:= \tbme\,(a+\Ld)^{-1}\,\bU\,(-b+\tLd)^{-1}\bme
\end{equation}
For these objects we have the following relations
\bse\label{eq:ssrels}\bea
&& 1+(p-a)s_{a,b}-(p-b)\wt{s}_{a,b}=\wt{v}_a w_b\  , \label{eq:ssrels_a} \\
&& (p+a+b)+(p^2+pb+b^2)s_{a,b}-(p^2+ap+a^2)\wt{s}_{a,b} = s_a \wt{w}_b-v_a \wt{t}_b+ pv_a \wt{w}_b\  ,
\label{eq:ssrels_b}
\eea\ese
as well as
\bse\label{eq:u01}\bea
&& \wt{u}_{0,1}+u_{1,0}=p\wt{u}_0-pu_0+\wt{u}_0 u_0\  , \\
&& \frac{p^3-q^3}{p-q+\wh{u}_0-\wt{u}_0} = p^2+pq+q^2 -(p+q)(\wh{\wt{u}}_0-u_0)-u_0\wh{\wt{u}}_0+\wh{\wt{u}}_{1,0}+u_{0,1}\   ,
\eea\ese
and
\bse\bea
\wt{s}_a &=& (p+u_0)\wt{v}_a-(p-a) v_a \  , \\
t_b &=& (p-b)\wt{w}_b-(p-\wt{u}_0)w_b\
\eea\ese
and
\bse\label{eq:zrrels}\bea
\wt{r}_a &=& p\wt{s}_a-(p-a) s_a+\wt{v}_a u_{0,1}\  , \\
z_b &=& (p-b) \wt{t}_b-p t_b+\wt{u}_{1,0} w_b\  ,
\eea\ese
and
\bse\label{eq:rzrels}\bea
r_a &=& (p^2+ap+a^2)\wt{v}_a-(p-\wt{u}_0)s_a-\left(p(p-\wt{u}_0)+\wt{u}_{1,0}\right) v_a\  , \\
\wt{z}_b &=& (p^2+pb+b^2) w_b+(p+u_0)\wt{t}_b-\left( p(p+u_0)+u_{0,1}\right) \wt{w}_b\  .
\eea\ese
All relations \eqref{eq:ssrels}-\eqref{eq:rzrels} also hold for their counterparts obtained by replacing the shift $\wt{\phantom{a}}$ by the shift $\wh{\phantom{a}}$
whilst replacing the parameter $p$ by $q$. We mention that from the relations \eqref{eq:u01} by eliminating the quantities $u_{0,1}$ and $u_{1,0}$ we obtain
immediately the lattice BSQ equation \eqref{eq:dBSQ} in the form:
\bea\label{eq:ddBSQ}
&& \frac{p^3-q^3}{p-q+\wh{\wh{u}}_0-\wh{\wt{u}}_0} - \frac{p^3-q^3}{p-q+\wh{\wt{u}}_0-\wt{\wt{u}}_0}= \nn \\
&& = (p-q+\wh{\wh{\wt{u}}}_0-\wh{\wt{\wt{u}}}_0)(2p+q+\wh{u}_0-\wh{\wh{\wt{\wt{u}}}}_0)-(p-q+\wh{u}_0-\wt{u}_0)(2p+q+u_0-\wh{\wt{\wt{u}}}_0)\  .
\eea

\subsection{Basic vectorial relations}
Introducing the 3-component vectors
\be\label{eq:vectors}
\bv_a:= \left( \begin{array}{c} \rho_a v_a \\ \rho_{\oa a} v_{\oa a} \\ \rho_{\oa^2 a} v_{\oa^2 a}\end{array}\right) \quad,\quad
\bw_b:= \left( \begin{array}{c} \sg_b w_b \\ \sg_{\oa b} w_{\oa b} \\ \sg_{\oa^2 b} w_{\oa^2 b}\end{array}\right)\  ,
\ee
where
\be\label{eq:rhosg}
\rho_a=\left(\frac{P_a}{p-a}\right)^n \left(\frac{Q_a}{q-a}\right)^m\quad,\quad
\sg_b=\left(\frac{p-b}{P_b}\right)^n \left(\frac{q-b}{Q_b}\right)^m
\ee
and similarly the vectors $\bs_a$, $\btt_b$, $\brr_a$ and $\bz_b$,
we obtain the sets of vectorial equations:
\bse\label{eq:st}\begin{eqnarray}
\wt{\bs}_a &=& (p+u_0)\wt{\bv}_a-P_a\bv_a  \label{eq:st_a} \\
\btt_b &=& P_b\wt{\bw}_b -(p-\wt{u}_0) \bw_b \label{eq:st_b}
\end{eqnarray}\ese
as well as
\bse\label{eq:rz}\begin{eqnarray}
\wt{\brr}_a &=& p\wt{\bs}_a-P_a\bs_a+u_{0,1}\wt{\bv}_a  \label{eq:rz_a} \\
\bz_b &=& P_b\wt{\btt}_b -p\btt_b+\wt{u}_{1,0} \bw_b\  .  \label{eq:rz_b}
\end{eqnarray}\ese
and
\bse\label{eq:zr}\begin{eqnarray}
\brr_a &=& P_a\wt{\bv}_a-(p-\wt{u}_0)\bs_a-\left(p(p-\wt{u}_0)+\wt{u}_{1,0}\right)\bv_a  \label{eq:zr_a} \\
\wt{\bz}_b &=& P_b \bw_b+(p+u_0)\wt{\btt}_b -\left(p(p+u_0)+u_{0,1}\right)\wt{\bw}_b\  .  \label{eq:zr_b}
\end{eqnarray}\ese

The main object of interest is the following quantity
\be\label{eq:main}
u:= \sum_{i,j=0}^2\,A_{ij} \rho_{\oa^i a}\sg_{\oa^j b}S(\oa^i a,\oa^j b)\quad {\rm where}\quad S(a,b)=s_{a,b}-\frac{1}{a-b}\  ,
\ee
for which, as a consequence of \eqref{eq:ssrels_a} and \eqref{eq:ssrels_b} respectively, we have the following basic relations
\bse\label{eq:uu}\begin{eqnarray}
P_a u-P_b \wt{u} &=& \wt{\bv}_a^t\,\bA\,\bw_b\   ,  \label{eq:uu_a} \\
P_b u-P_a \wt{u} &=& \bs_a^t\,\bA\,\wt{\bw}_b-\bv_a^t\,\bA\,\wt{\btt}_b +p\bv_a^t\,\bA\,\wt{\bw}_b    \   . \label{eq:uu_b}
\end{eqnarray}\ese
It is in terms of the quantity $u$ that we will obtain the higher-rank (BSQ) analogue of the Q3 equation. We will now show how the equation
for this object is constructed.

Introducing the quantities:
\be\label{eq:UVWZ}
U:=\bv_a^t\,\bA\,\bw_b\quad,\quad V:=\bv_a^t\,\bA\,\btt_b\quad,\quad W:=\bs_a^t\,\bA\,\bw_b\quad,\quad Z:=\bs_a^t\,\bA\,\btt_b\  ,
\ee
and setting ~$P=P_aP_b$~,~$Q=Q_aQ_b$~, we have the following relations
\bse\label{eq:Miuras}\bea
P\wt{u}-Q\wh{u}-(p^3-q^3)u &=& -(p-q+\wh{u}_0-\wt{u}_0)\left[ (p+q-\wh{\wt{u}}_0) U+W\right] \  , \label{eq:UW} \\
P\wh{u}-Q\wt{u}-(p^3-q^3)\wh{\wt{u}} &=& (p-q+\wh{u}_0-\wt{u}_0)\left[ (p+q+u_0) \wh{\wt{U}}-\wh{\wt{V}}\right] \  , \label{eq:UV} \\
Pu-Q\wh{\wt{u}}-(p^3-q^3)\wt{u} &=& \wt{Z}-(p+u_0)\wt{V}+(q-\wh{\wt{u}}_0)\wt{W} \nn \\
&& \quad + \left[p(p+u_0)+q(q-\wh{\wt{u}}_0)+\wh{\wt{u}}_{1,0}+u_{0,1}\right]\wt{U}\  , \label{eq:UZ}
\eea\ese
which we will think of as Miura type relations. Similar relations appeared in \cite{NAH} connecting solutions of Q3 to solutions of H1 
(the potential KdV equation), which is the $N=2$ analogue of the lattice BSQ equation \eqref{eq:ddBSQ}. 

Several additional relations can be derived by combining the relations \eqref{eq:st}, \eqref{eq:rz} and \eqref{eq:zr}, namely from
\eqref{eq:st} by elimination of $\bs$ respectively $\btt$ we get:
\bse\label{eq:vw}\bea
P_a \wh{\bv}_a-Q_a \wt{\bv}_a &=& (p-q+\wh{u}_0-\wt{u}_0) \wh{\wt{\bv}}_a\  , \\
P_b \wt{\bw}_b-Q_b \wh{\bw}_b &=& (p-q+\wh{u}_0-\wt{u}_0) \bw_b\  .
\eea\ese
Similarly, by combining \eqref{eq:rz} in two different lattice directions we find
\bse\label{eq:wv}\bea
P_a \wt{\bv}_a-Q_a \wh{\bv}_a &=& (p-q+\wh{u}_0-\wt{u}_0)\left[ \bs_a+ (p+q-\wh{\wt{u}}_0)\bv_a\right]\  , \\
P_b \wh{\bw}_b-Q_b \wt{\bw}_b &=& -(p-q+\wh{u}_0-\wt{u}_0)\left[\, \wh{\wt{\btt}}_b-(p+q+u_0)\wh{\wt{\bw}}_b\right]\  .
\eea\ese
Furthermore, by combining \eqref{eq:zr} in two different lattice directions we find
\bse\label{eq:ts}\bea
P_a \wh{\bs}_a-Q_a \wt{\bs}_a &=& (p-q)\wh{\wt{\bs}}_a+ \left[ (p-q+\wh{u}_0-\wt{u}_0)u_0 -p\wt{u}_0+q\wh{u}_0 \right] \wh{\wt{\bv}}_a\  , \\
P_b \wt{\btt}_b-Q_b \wh{\btt}_b &=& (p-q)\btt_b+\left[(p-q+\wh{u}_0-\wt{u}_0)\wh{\wt{u}}_0-p\wh{u}_0+q\wt{u}_0 \right] \bw_b\  .
\eea\ese
We will now proceed by using these constitutive relations to derive the lattice equations, involving primarily three objects, $u$, $U$ and $u_0$.

\section{Lattice BSQ analogue of Q3}
\setcounter{equation}{0}

The main object in the construction, following the relations obtained in the previous section, is the following 3$\times$3 matrix
\be\label{eq:E}
E_{a,b}:= \left( \begin{array}{ccccc}
\bv_a^t\,\bA\,\bw_b &,& \bv_a^t\,\bA\,\btt_b &,& \bv_a^t\,\bA\,\bz_b \\
\bs_a^t\,\bA\,\bw_b &,& \bs_a^t\,\bA\,\btt_b &,& \bs_a^t\,\bA\,\bz_b \\
\brr_a^t\,\bA\,\bw_b &,& \brr_a^t\,\bA\,\btt_b &,& \brr_a^t\,\bA\,\bz_b
\end{array}\right)\  .
\ee
The determinant of this matrix can be expressed in two different ways. On the one hand, by applying a special case of the
well-known Weinstein-Aronszajn formula, we have
\begin{eqnarray}\label{eq:detE} 
\det(E_{a,b}) &=& \det(\bA)\,\det(\bv_a,\bs_a,\brr_a)\,\det(\bw_b,\btt_b,\bz_b)  \nn \\
              &=& \left(\frac{P_a}{P_b}\right)^n \left(\frac{Q_a}{Q_b}\right)^m \alpha\beta \det(\bA)\  ,
\end{eqnarray}
in which $\alpha$ and $\beta$ are constants (w.r.t. the variables $n$ and $m$) but will depend on the parameters $a$ and $b$ respectively.
In fact, by applying \eqref{eq:rz_a} and subsequently \eqref{eq:st_a} and \eqref{eq:zr_a} we have
\begin{eqnarray*}
&& \det(\wt{\bv}_a,\wt{\bs}_a,\wt{\brr}_a)= -P_a \det(\wt{\bv}_a,\wt{\bs}_a,\bs_a)= P_a^2 \det(\wt{\bv}_a,\bv_a,\bs_a)=P_a\det(\brr_a,\bv_a,\bs_a) \\
&& \Rightarrow \quad \wt{\det(\bv_a,\bs_a,\brr_a)}=P_a \det(\bv_a,\bs_a,\brr_a)\  ,
\end{eqnarray*}
similarly we find in the other lattice direction ~$\wh{\det(\bv_a,\bs_a,\brr_a)}=Q_a \det(\bv_a,\bs_a,\brr_a)$~. Thus, we have
\[ \det(\bv_a,\bs_a,\brr_a)=P_a^n Q_a^m \det(\bv_a,\bs_a,\brr_a)|_{n=m=0}=:P_a^nQ_a^m \alpha\   , \]
setting the value of this determinant at $n=m=0$ equal to $\alpha$. By a similar argument we find that
determinant ~$\det(\bw_b,\btt_b,\bz_b)$~ can be found to be equal to
\[ \det(\bw_b,\btt_b,\bz_b)=P_b^{-n} Q_b^{-m} \det(\bw_b,\btt_b,\bz_b)|_{n=m=0}=:P_b^{-n} Q_b^{-m} \beta\   , \]
setting the value of the latter determinant at $n=m=0$ equal to $\beta$.

On the other hand, using the relations \eqref{eq:st} \eqref{eq:rz} we can rewrite the determinant of $E_{a,b}$ as
\be\label{eq:EE}
\det(E_{a,b})=-PQ \left| \begin{array}{ccccc}
U &,& \bv_a^t\,\bA\,\wh{\bw}_b &,& \bv_a^t\,\bA\,{\hypotilde 0 \bw}_b \\
{\hypotilde 0 \bv}_a^t\,\bA\,\bw_b &,& {\hypotilde 0 \bv}_a^t\,\bA\,\wh{\bw}_b &,& \ut{U} \\
\wh{\bv}_a^t\,\bA\,\bw_b &,& \wh{U} &,& \wh{\bv}_a^t\,\bA\,{\hypotilde 0 \bw}_b
\end{array}\right|\  .
\ee
(the under-accents ~$\underaccent{\wtilde}{\cdot}$~ and ~$\underaccent{\what}{\cdot}$~ denoting the shifts in the opposite directions to the 
shifts ~$\wt{\cdot}$~ and ~$\wh{\cdot}$~), where the entries can be evaluated using the relations
\bse\label{eq:vwrels}\bea
{\wh{\wt{\bv}}_a}^t\,\bA\,\bw_b &=& \frac{Q_a P_b\wt{u}-P_a Q_b\wh{u}}{p-q+\wh{u}_0-\wt{u}_0}\   , \label{eq:vwrels_a} \\
P_a {\hypotilde 0 \bv}_a^t\,\bA\,\bw_b &=& (2p+q+{\hypotilde 0 u}_0-\wh{\wt{u}}_0)U +
\frac{P\wt{u}-Q\wh{u}-(p^3-q^3)u}{p-q+\wh{u}_0-\wt{u}_0} \  , \label{eq:vwrels_b}\\
P_b \bv_a^t\,\bA\,\wt{\bw}_b &=& (2p+q+\underaccent{\what}{\underaccent{\wtilde}{u}}_0-\wt{u}_0)U -
\frac{P{\hypotilde 0 u}-Q{\hypohat 0 u}-(p^3-q^3)u}{p-q+{\hypotilde 0 u}_0-{\hypohat 0 u}_0} \  , \label{eq:vwrels_c}\\
P_b Q_b \bv_a^t\,\bA\,\wh{\wt{\bw}}_b &=& \frac{P\wt{U}-Q\wh{U}}{p-q+\wh{u}_0-\wt{u}_0} +P\wt{u}+Q\wh{u}-
(p^3+q^3-a^3-b^3)u\   , \label{eq:vwrels_d}\\
P_a Q_b \bv_a^t\,\bA\,\wh{\wt{\bw}}_b &=& \frac{p^3-q^3}{p-q+\wh{u}_0-\wt{u}_0}\,\wt{U}-Pu+Q\wh{\wt{u}}+
(p^3-q^3)\wt{u}\   ,\label{eq:vwrels_e}
\eea\ese
as well as, where relevant, their counterparts obtained by interchanging the shifts $\wt{\phantom{a}}$ and $\wh{\phantom{a}}$ and
simultaneously interchanging $p$ and $q$.
Eqs. \eqref{eq:vwrels} can be deduced from \eqref{eq:Miuras} and \eqref{eq:st} through the definitions \eqref{eq:UVWZ}, using also 
\eqref{eq:uu} and the relations \eqref{eq:vw}-\eqref{eq:ts}. In addition to these relations, we also obtain the following important relation
\begin{equation}\label{eq:PQUU}
\frac{P_bQ_a \wt{U}-Q_bP_a\wh{U}}{p-q+\wh{u}_0-\wt{u}_0} =P_aQ_a u+P_bQ_b\wh{\wt{u}}-Q_aP_b\wt{u}-P_aQ_b\wh{u}\  ,
\end{equation}
accounting for the internal consistency of the relations \eqref{eq:vwrels}. Eq. \eqref{eq:PQUU} can be used to eliminate 
the canonical combination ~$\gm:=p-q+\wh{u}_0-\wt{u}_0$~ from all relations \eqref{eq:vwrels} in favour of the variables $U$ and $u$. 
Let us now formulate the main statement:\vspace{.2cm}

\noindent
{\bf Proposition:} {\it Eq. \eqref{eq:detE}, where $\det(E_{a,b})$ can be rewritten in various forms such as the one exhibited in 
\eqref{eq:EE} constitutes the $N=3$ higher rank generalization of the Q3 equation, where the entries of the determinant can be written 
in terms of the ``conjugate'' variables $u$ and $U$ using the relations in \eqref{eq:vwrels} (and their counterparts, where relevant, 
interchanging $p$ and $q$ and ~$\wt{\phantom{a}}$~ and ~$\wh{\phantom{a}}$~). The system is supplemented by a set of linear ordinary 
difference equations, in each of the lattice directions, for $U$ obtained by eliminating between \eqref{eq:vwrels_b} and \eqref{eq:vwrels_c} 
(and their counterparts in the other lattice direction) the left-hand sides. In all expressions the canonical quantity $\gm$ is obtained from 
\eqref{eq:PQUU} in terms of $u$ and $U$.} \vspace{.2cm} 

\paragraph{} We will refrain from writing out the resulting system in explicit form, which may not be insightful. The presentation 
of the system may depend, in fact, on the choice of the combination of dependent variables considered as the prime variables. For instance, $U$ can 
actually be solved from a cubic equation which is obtained from $\det(E_{a,b})$ by using the linear difference equations for $U$ (alluded to in the proposition), 
to eliminate $\wh{U}$ and ${\hypotilde 0 U}$ from \eqref{eq:EE}. We intend, in a future publication, to address these options in more detail in order 
to find a convenient representation of this system which is incorporated in \eqref{eq:detE} and the relations \eqref{eq:vwrels}, 
\eqref{eq:PQUU}.

\section{Lax representation}
\setcounter{equation}{0}

The Lax representations can be obtained in a systematic way from the relations \eqref{eq:ukrels}, i.e. in terms of quantities
defined on the basis of the vectors $\bu_k$ yielding the objects obeying the relevant linear systems of equations.  The first step is to define
the quantities
\be\label{eq:uka}
u_k^{(j)}:= \tbme\,\Ld^j\,\bu_k\quad,\quad u_k(a);=\tbme\,(a+\Ld)^{-1}\,\bu_k\  ,
\ee
for which we can derive directly from \eqref{eq:ukrels} the following relations
\bse\label{eq:eukrels}\bea
\wt{u}_k^{(0)} &=& (p-\wt{u}_0)u_k^{(0)}+u_k^{(1)}\   , \\
\wt{u}_k(a) &=& (p-a) u_k(a)+\wt{v}_a u_k^{(0)}\   ,  \\
(p^3+k^3) u_k(a) &=& (p^2+ap+a^2) \wt{u}_k(a) -(p v_a+s_a) \wt{u}_k^{(0)}+v_a \wt{u}_k^{(1)}\  ,
\eea\ese
and a set of similar relations with $p$ replaced by $q$ and $\wt{\phantom{a}}$ by $\wh{\phantom{a}}$. The next step is to introduce the
3-component vector
\be\label{eq:uka}
\bu_k(a)=\left( \begin{array}{c} \rho_a u_k(a) \\ \rho_{\oa a} u_k(\oa a) \\ \rho_{\oa^2 a} u_k(\oa^2 a)\end{array}\right)\  ,
\ee
with the plane-wave factors $\rho_a$ as given in \eqref{eq:rhosg}. With the definitions of the 3-vectors \eqref{eq:vectors} we now
obtain from \eqref{eq:eukrels} the following relations
\bse\label{eq:3vecrels}\bea
\wt{\bu}_k(a) &=& P_a \bu_k(a)+\wt{\bv}_a\,u_k^{(0)}\   , \label{eq:3vecs_a} \\
(p^3+k^3) \bu_k(a) &=& P_a \wt{\bu}_k(a)-(p\bv_a+\bs_a) \wt{u}_k^{(0)}+\bv_a \wt{u}_k^{(1)}\  , \label{eq:3vecs_b}
\eea \ese
with similar ones involving the other shift of the lattice.

The third step is to introduce the scalar object
\[ u_k(a,b):= \bw_b^t\,\bA^t\,\bu_k(a)\  , \]
where $\bA^t$ is the transpose of the 3$\times$3 matrix $\bA$ used in \eqref{eq:UVWZ}. Multiplying the relations \eqref{eq:3vecrels} from the
left by $\bw_b^t$ and $\wt{\bw}_b^t$ respectively, we obtain the set
\begin{eqnarray*}
\bw_b^t\,\bA^t\,\wt{\bu}_k(a) &=& P_a u_k(a,b) + (\bw_b^t\,\bA^t\,\wt{\bv}_a)\,u_k^{(0)}\  , \\
\wt{u}_k(a,b) &=& P_a \wt{\bw}_b\,\bA^t\,\bu_k(a) + \wt{U} u_k^{(0)}\  , \\
(p^3+k^3) u_k(a,b) &=& P_a\,\bw_b^t\,\bA^t\,\wt{\bu}_k(a)-(pU+W) \wt{u}_k^{(0)}+U \wt{u}_k^{(1)}\  , \\
(p^3+k^3) \wt{\bw}_b^t\,\bA^t\,\bu_k(a) &=& P_a \wt{u}_k(a,b)-\left( p\wt{\bw}_b^t\,\bA^t\,\bv_a + \wt{\bw}_b^t\,\bA^t\,\bs_a\right)\wt{u}_k^{(0)}
+(\wt{\bw}_b^t\,\bA^t\,\bv_a) \wt{u}_k^{(1)}\  .
\end{eqnarray*}
Upon elimination from this set the quantities ~$\bw_b^t\,\bA^t\,\wt{\bu}_k(a)$~ and ~$\wt{\bw}_b^t\,\bA^t\,\bu_k(a)$~ we obtain the following relations
\bse\bea
(a^3+k^3) u_k(a,b) &=& P_a (\bw_b^t\,\bA^t\,\wt{\bv}_a)u_k^{(0)} -(pU+W) \wt{u}_k^{(0)}+ U \wt{u}_k^{(1)}\   , \\
(a^3+k^3) \wt{u}_k(a,b) &=& (p^3+k^3) \wt{U} u_k^{(0)} -P_a\left( p\wt{\bw}_b^t\,\bA^t\,\bv_a + \wt{\bw}_b^t\,\bA^t\,\bs_a\right)\wt{u}_k^{(0)}
+P_a(\wt{\bw}_b^t\,\bA^t\,\bv_a) \wt{u}_k^{(1)}\  . \nn  \\
&&
\eea\ese
Introducing, finally, the 3-component vector
\[ \bphi_k(a,b):= \left(\begin{array}{c}
u_k^{(0)} \\ u_k^{(1)} \\ u_k(a,b)\end{array}\right) \]
we now obtain the following set of linear equations (Lax pair) for $\bphi_k(a,b)$:
\bse\label{eq:Lax}\be
\wt{\bphi}_k(a,b)=\bL_k(a,b)\,\bphi_k(a,b)\quad,\quad  \wh{\bphi}_k(a,b)=\bM_k(a,b)\,\bphi_k(a,b)
\ee
with the Lax matrices $\bL_k$ and $\bM_k$. $\bL_k$ is given by
\be\label{eq:Laxmat}
\bL_k(a,b)=\frac{1}{U}\left(\begin{array}{ccccc}
(p-\wt{u}_0)U &,& U &,& 0 \\
(p-\wt{u}_0)(pU+W)-P_a E &,& pU+W &,& a^3+k^3 \\
\ast &,& \frac{1}{a^3+k^3}P_a(\Gamma W-U\Delta) &,& P_a\Gamma
\end{array}\right)\  ,
\ee\ese
in which the espression $\ast$ stands for
\[ \ast=U\wt{U}+\frac{1}{a^3+k^3}\left[(p^3-a^3)(U\wt{U}-E\Gamma)+P_a(p-\wt{u}_0)(\Gamma W-U\Delta)\right] \  , \]
i.e., such that the determinant is given by ~$\det(L_k(a,b))=(p^3+k^3)\wt{U}/U$~. Here we have used the following abbreviations:
\[
E:= \bw_b^t\,\bA^t\,\wt{\bv}_a\quad,\quad \Gamma:=\wt{\bw}_b^t\,\bA^t\,\bv_a\quad,\quad \Delta:=\wt{\bw}_b^t\,\bA^t\,\bs_a\  , \]
which can be identified with the objects given in \eqref{eq:vwrels} leading to
\be
E=P_a u-P_b\wt{u}\quad,\quad P_a\Gamma=(p+u_0)\wt{U}-\wt{W}\quad.\quad P_a\Delta=p\wt{W}+u_{0,1}\wt{U}-\wt{R}\  ,
\ee
where we have set ~$R:=\brr^t_a \bA\bw_b$~. The latter quantity can be expressed in terms of more familiar objects, 
such as 
\[ R=(p^3-a^3)u-P\wt{u}-(p-\wt{u}_0)W-\left( p(p-\wt{u}_0)+\wt{u}_{1,0}\right) U=pW+{\hypotilde 0 u}_{0,1}U\   , \] 

The matrix $\bM_k$ is given by similar expressions making the usual replacements, and the compatibility condition of the Lax pair, i.e., 
the Lax equations $\wh{\bL}_k\bM_k=\wt{\bM}_k\bL_k$, can be explicitly verified entry-by-entry and leads to a set of relations 
compatible to the relations \eqref{eq:vwrels}. Once again, for the system the question arises as to what is the best realisation of the 
system, in order to obtain an \textit{effective} Lax pair which could be used, e.g. for the construction of wide classes of solutions 
of the system.

\section{Conclusions}

In this paper we have derived a new integrable lattice system of higher rank which incorporates the special subcases of known BSQ type solutions 
as were presented in \cite{GD}. In fact, by choosing the arbitrary matrix $\bA$ to have a special form, such that the inner products of the type 
$\mbx^t\bA\,\mby$ factorize, it can be shown that the system reduces again to equations of the form \eqref{eq:interBSQ}. A new feature is that  
whenever $\det(\bA)\neq 0$ the analogous term associated with the $\dd^2$-parameter (in the form of Q3 as given in the ABS list of \cite{ABS}) 
becomes non-autonomous. This is a feature that we expect to persist in the higher members of the Gel'fand-Dikii  
hierarchy as well, and only if $N=2$ this ``deformation term'' of the equation is independent of the variables $n$ and $m$. 
We stopped short of writing the lattice system in fully explicit form for reasons explained earlier, and finding a convenient explicit realisation 
would be a next step. Many questions are obviously worth considering, such as the construction of elliptic solutions along the line of \cite{NA} 
(which I expect would lead to basically the same system of equations), the question about the role of (in this case $SL(3)$-) invariant polynomials 
associated with the system, the related question regarding Lagrangian structures (which exists for the all the lattice Gel'fand-Dikii hierarchy, cf. 
\cite{LN}), and the question about a possible existence of elliptic variants of 
the system. Since the lattice BSQ equations such as \eqref{eq:dBSQ}, can sometimes exhibit interesting geometrical aspects, e.g. the emergence of 
the pentagram map, cf. \cite{OST}, matters regarding the discrete differential geometry in the sense of \cite{BS2} may be worth considering 
as well.  These and other issues will be relegated to future work.

\section*{Acknowledgements}
This work was finalised while commencing a visit at Shanghai University. The author is grateful to Prof. D-J Zhang for his hospitality.

\end{document}